# Measurement of the Free Neutron Lifetime in a Magneto-Gravitational Trap with *In Situ* Detection.

R. Musedinovic[1], L. S. Blokland[2], C. B. Cude-Woods[3], M. Singh[3], M. A. Blatnik[3,4], N. Callahan[5], J. H. Choi[1], S. Clayton[3,*], B. W. Filippone[4], W. R. Fox[2], E. Fries[4], P. Geltenbort[6], F. M. Gonzalez[7], L. Hayen[1], K. P. Hickerson[4], A. T. Holley[8], T. M. Ito[3], A. Komives[9], S Lin[3], Chen-Yu Liu[10], M. F. Makela[3], C. M. O'Shaughnessy[3], R. W. Pattie Jr.[11], J. C. Ramsey[7], D. J. Salvat[2], A. Saunders[7], S. J. Seestrom[3], E. I. Sharapov[12], Z. Tang[3], F. W. Uhrich[3], J. Vanderwerp[2], P. Walstrom[3], Z. Wang[3], A. R. Young[1,13], and C. L. Morris[3]

[1]Department of Physics, North Carolina State University, Raleigh, NC 27695, USA

[2]Department of Physics, Indiana University, Bloomington, IN, 47405, USA

[3]Los Alamos National Laboratory, Los Alamos, NM, USA, 87545, USA

[4]Kellogg Radiation Laboratory, California Institute of Technology, Pasadena, CA 91125, USA

[5]Argonne National Laboratory, Lemont, IL 60439, USA

[6]Institut Laue-Langevin, CS 20156, 38042 Grenoble Cedex 9, France

[7]Oak Ridge National Laboratory, Oak Ridge, TN 37831, USA

[8]Tennessee Technological University, Cookeville, TN 38505, USA

[9]DePauw University, Greencastle, IN 46135, USA

[10]University of Illinois, Urbana, IL 61801, USA

[11]East Tennessee State University, Johnson City, TN 37614, USA

[12]Joint Institute for Nuclear Research, 141980 Dubna, Russia

[13]Triangle Universities Nuclear Laboratory, Durham, NC 27708, USA

*Corresponding author: S. Clayton, sclayton@lanl.gov

**Abstract** Here we publish three years of data for the UCNt experiment performed at the Los Alamos Ultra Cold Neutron Facility at the Los Alamos Neutron Science Center. These data are in addition to our previously published data. Our goals in this paper are to better understand and quantify systematic uncertainties and to improve the lifetime statistical precision. We report a measured value for these runs from 2020-2022 for the neutron lifetime of 877.94±0.37 s; when all the data from UCNτ are averaged we report an updated value for the lifetime of 877.82±0.22 (statistical)+0.20-0.17 (systematic) s. We utilized improved monitor detectors, reduced our correction due to UCN upscattering on ambient gas, and employed four different main UCN detector geometries both to reduce the correction required for rate dependence and explore potential contributions due to phase space evolution.

## Introduction

The decay of the free neutron $n \rightarrow p + e^- + \bar{\nu}_e$ is the simplest example of nuclear $\beta$-decay, and measurements of decay observables have implications for the Standard Model of particle physics and cosmology. The mean neutron lifetime, $\tau_n$, is needed as an input to predict primordial light element abundances[1]. The combination of the lifetime and neutron decay correlation parameters test the $V$-$A$ structure of the weak interaction without complications from nuclear structure corrections[2] Recent and forthcoming neutron $\beta$-decay experiments can be used to extract the magnitude of the Cabibbo-Kobayashi-Maskawa (CKM) matrix element $V_{ud}$ with a precision approaching that from studies of super-allowed $0^+ \rightarrow 0^+$ nuclear $\beta$-decays. Further, these tests can probe the existence of beyond-standard-model interactions that could evade detection in high-energy collider experiments [3, 4].

The history of $\tau_n$ measurements, as well as the value evaluated by the Particle Data Group (PDG)[5], are plotted in Figure 1. The present work reports additional results of an ongoing experiment[6] to measure $\tau_n$ with smaller systematic corrections than previous efforts and using a blinded analysis to avoid confirmation bias. The new data reported here were acquired in 2020, 2021, and 2022 at the Los Alamos Ultra-Cold Neutron Facility[7, 8] at the Los Alamos Neutron Science Center (LANSCE).

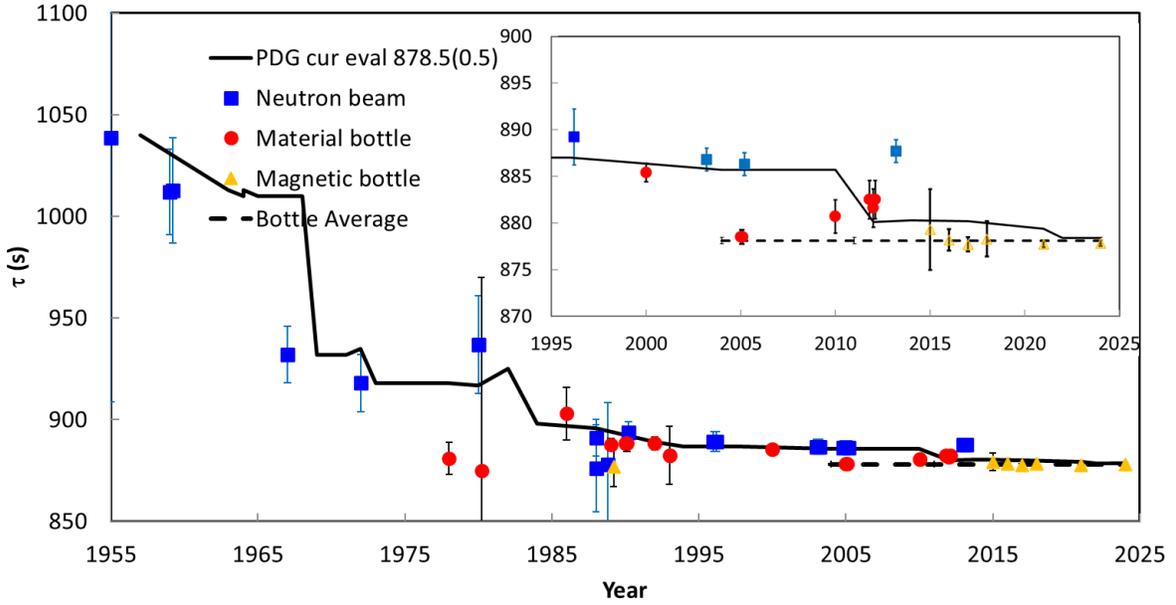

Figure 1) The history of neutron lifetime measurements and the evaluated neutron lifetime[5, 9]. The inset shows the most recent results on an expanded scale.

In the standard model, the following relationship holds[10]:

$$\tau_n^{-1} = \frac{m_e^5}{2\pi^3} G_F^2 |V_{ud}|^2 \left(1 + 3g_A^2\right)(1 + RC)\, f\,, \qquad (1)$$

where $G_F$ is the Fermi coupling constant, $g_A$ is the neutron weak axial-current coupling, $f$ is a decay phase-space factor, and $RC$ represents the electroweak radiative corrections. The "inner" radiative correction contains model-dependent hadronic structure and short-distance QCD physics and is the dominant source of theoretical uncertainty in Equation 1[10, 11].

Steady refinement of the theoretical analysis of the Standard Model expectations[11-17] for charged current decays of kaons, neutrons, and nuclei has produced strong evidence for discrepancies between the decay observables and standard model expectations collectively referred to as the Cabibbo Angle Anomaly (CAA)[18, 19]. In particular, violation of the expected unitarity of the top row of the Cabibbo-Kobayashi-Maskawa matrix appear at roughly the 3σ level or more in these data. Analysis of the CAA in a model-independent framework using effective field theory[20, 21] now incorporates low energy observables such as the neutron lifetime, electroweak precision observables from the Large Electron-Positron collider (LEP), and the recent measurement of the W mass[22], in addition to constraints from the Large Hadron Collider. With this approach, Reference 20 identifies potential new physics originating from right-handed currents[20] or vector-like quarks[23]. A number of other scenarios have been studied. Reference[24], for example, presents analysis of the CAA incorporating data from the decay of bottom-quarks and the muon anomalous magnetic moment that points to new physics from leptoquarks[25].

Discriminating between these model scenarios requires significant improvement in the current experimental observables. Neutron $\beta$-decay data can play a crucial role in these analyses by reducing uncertainties in the input value for the $V_{ud}$ parameter due to nuclear structure uncertainties associated with superallowed nuclear beta decays[17, 26]. This theoretical work increases the motivation for improving the accuracy in both neutron lifetime and beta decay correlations.

Measurements of $\tau_n$ are generally performed using either the so-called "beam" or "bottle" technique. The beam technique consists of passing a slow neutron beam through a decay volume of known length and cross-sectional area, and counting neutron decay products ($e^-$, p, or both) within that volume. With an absolute measurement of the neutron flux and absolute determination of the detector efficiencies, the partial neutron decay rate for neutrons that produce the detected decay product in the final state can be determined. By far the most precise of these beam experiments uses a quasi-Penning trap and silicon surface-barrier detector to count protons from a cold neutron beam at the National Institute of Standards and Technology, NIST[27]. The most recent evaluation gives $\tau_n =$ 887.7±1.2[stat]±1.9[syst] s[28].

The "bottle" technique consists of introducing ultracold neutrons (UCN) with kinetic energy $E \leq 100$ neV into a material or magnetic bottle, storing the UCN for varying times, and counting the surviving neutrons to determine the storage lifetime. The most precise measurements using material bottles differ from the NIST experiment by as much as 4.4-$\sigma$[29-33]. A measurement using a cylindrical magnetic bottle at the Institut Laue-Langevin (ILL)[34] and measurements with the UCN$\tau$ apparatus, an asymmetric bowl-shaped magnetic trap at Los Alamos National Laboratory (LANL), give consistent results which also disagree with the NIST experiment by more than 4-$\sigma$.

This "neutron lifetime puzzle[35]" indicates either the existence of new physics leading to a decay channel without protons in the final state, or the presence of inadequately assessed or unidentified systematic effects in at least one of the experimental techniques. The former could be induced by the decay of neutrons to dark-matter particles[36], but such decay channels are constrained by the properties of neutron stars[37-39] and by direct searches looking for specific decay signatures [40-42]. The latter indicates a need for new experimental techniques to complement existing approaches and mitigate potential systematic effects.

In the present work, we provide an update of a $\tau_n$ measurement first reported in reference [43] using the UCN$\tau$ apparatus. The apparatus eliminates losses associated with material UCN bottles and utilizes novel detector technology to develop data-driven assessments of potential systematic effects[44][45]. The analysis reported here is aimed at developing new methods to characterize and reduce systematic effects well below 0.2 s, needed for the upcoming experiment, UCN$\tau$+, that will use a new elevator loading technique to increase the number of loaded UCN by a factor of 5-10.

## Experimental Configurations

The experimental configuration used for this experiment, shown in Figure 2, was similar to that in Ref. 6. We used a number of monitors to measure and study the UCN fluence; all of the monitor detectors were $^{10}$B coated ZnS:Ag UCN detectors[46]. As in previous years, a UCN conditioning volume (Round House or RH) was used to minimize the impact of fast variations in UCN production on UCN loading, especially near the end of the loading period. A significant improvement in normalization was achieved by emptying and counting the UCN remaining in that volume at the end of the fill using the RH Dump Detector. One monitor was upstream of the RH (Gate Valve Monitor) and one was mounted above the trappable UCN height in the RH (Active Cleaner). We also developed and used several versions of the primary UCN detector (UCN Dagger Detector) that were lowered into the active volume of the trap to measure surviving neutrons[47].

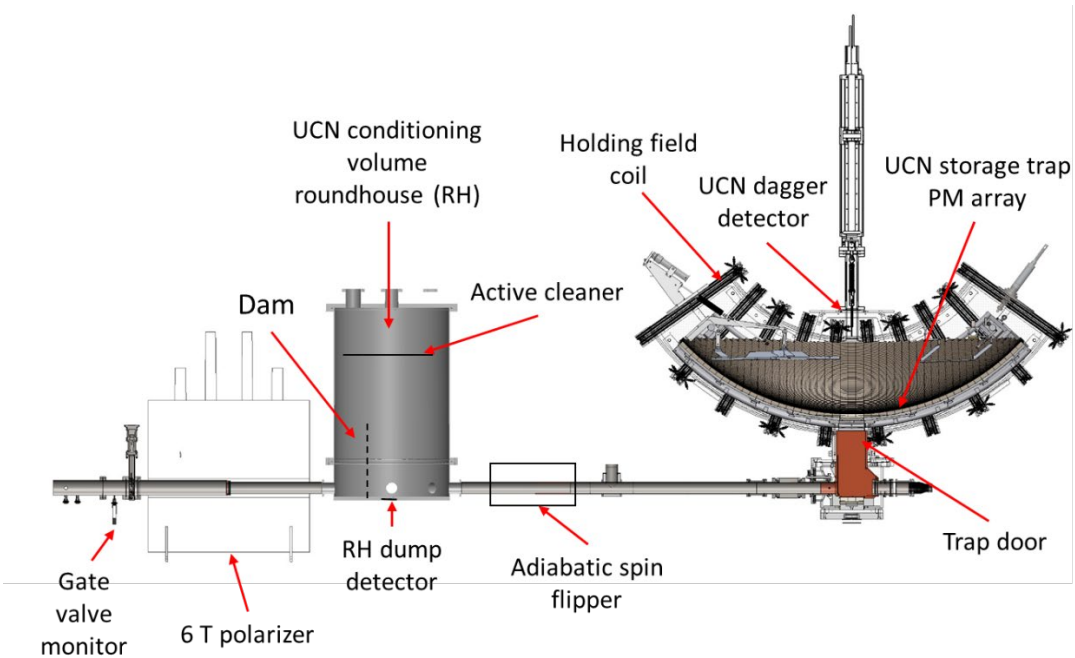

Figure 2) Experimental layout showing the location of the major components of this experiment. The UCN source is off the picture to the left. The RH dump detector is the primary normalization counter for these results. The dam was only used for selected runs and for testing.

In the experiment, each run is characterized by a storage time $T_{store}$. Neutrons are produced by 800 MeV protons delivered to a tungsten spallation target in ~0.5 s pulse strings every 5 s during the loading time of 300 s. During this time, a trap door at the bottom of the trap is lowered, and UCN are loaded into the trap. A cleaner to absorb high energy UCN is lowered into the top of the trap to a height of 38 cm from the bottom during the loading time and remains in the trap for an additional cleaning time (typically 50 s) after the trap door is closed. The dagger detector is partially lowered into the top of the trap to help with cleaning; the bottom edge is positioned at the same height as the cleaner at the start of filling and raised above the trap during the storage time, which begins when the cleaner and dagger are raised out of the trap. After the storage time, the dagger detector is typically first lowered to the cleaning position (to search for uncleaned or heated neutrons) and finally lowered to 1cm from the bottom of the trap where it counts stored neutrons for 300 s. These parameters are variable and were changed to perform systematic studies.

We made changes to the dagger throughout the experimental campaigns aimed at both increasing the efficiency of the detector and reducing peak count rates. The dagger consists of $^{10}$B-coated-ZnS scintillator laminated to an acrylic plate. Wavelength shifting fibers conduct the scintillation light induced by charged particles from the reaction $^{10}$B(n,$\alpha$)$^{7}$Li into pairs of phototubes. A photograph of the dagger used in 2022 is shown in Figure 3.

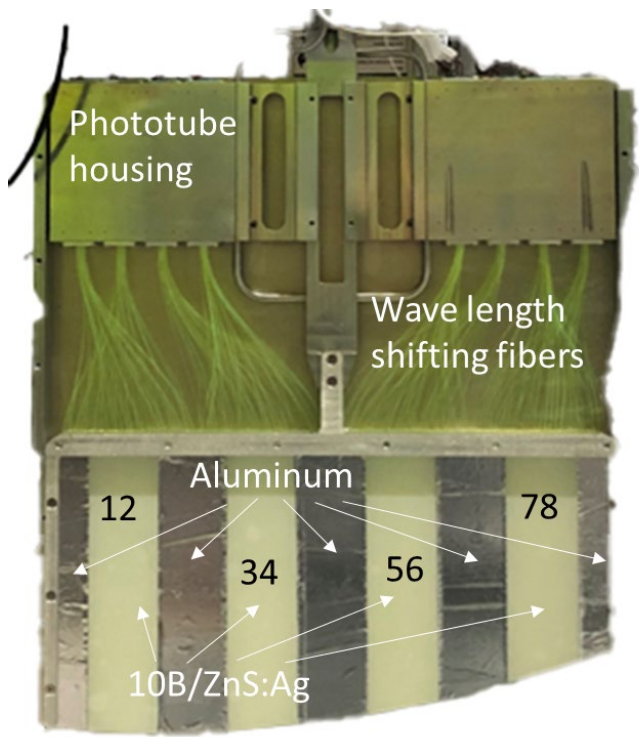

Figure 3) Photograph of the high counting rate dagger described in the text. An aluminum plate that covered the exposed wavelength shifting fibers during data acquisition has been removed for clarity. The strip numbers are labeled in black.

The small negative Fermi potential of $^{10}$B, combined with its high neutron absorption cross section ensures high efficiency for UCN counting. Different thicknesses of $^{10}$B, varying from 2 nm to 120 nm, have been tested and found to give statistically consistent lifetimes but very different counting times. Improved coating techniques produced transparent 120 nm $^{10}$B coatings used in the 2021 and 2022 daggers. A low-rate version of the dagger was tested in 2021 by covering half (one complete side) of the detector with UCN reflective aluminum. In 2020 and 2021 the light was distributed into two phototubes each covering the full dagger. In 2022 the dagger was divided into four vertical sections and utilized eight phototubes, with a pair of phototubes collecting light from each vertical section. This change reduced the maximum count rate in any section by about 4 from the previous dagger. The counting rate was further reduced by replacing part of the active area with aluminum strips (Figure 3). Unloading time distributions characterize the detected UCN as a function of time. Histograms depicting the counted UCN as a function of time after dagger motion started for each of these daggers are shown in Figure 4.

We studied potential systematic effects such as background, UCN loading, uncleaned neutrons, and UCN heating. Monte Carlo-generated pseudo data were used to test different coincidence algorithms and rate-dependent corrections to the UCN counts measured (see analysis section). We also tested inserting variable height barriers (dams) into the Roundhouse (see Figure 2) for the purpose of increasing the fidelity of the Roundhouse in representing the UCN spectrum ultimately stored in the UCN trap.

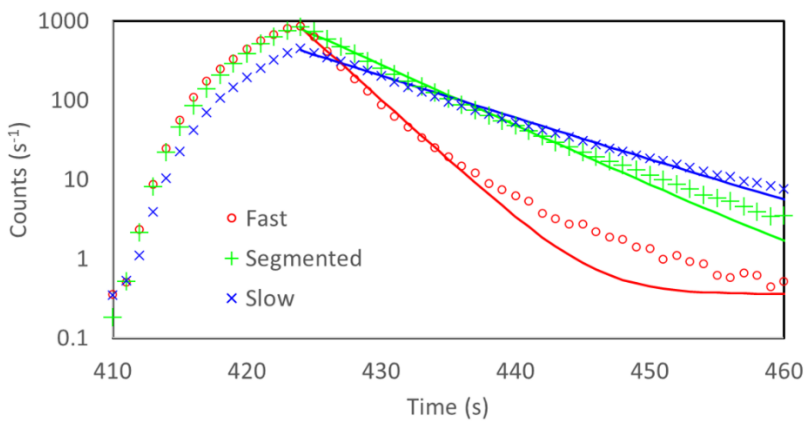

Figure 4) Unloading curve for each of the daggers described in the text. The data are an average of all the 20 s holding time counting curves for each dagger. The solid curves are single exponential plus constant background fits to the data.

## Analysis

The approach to analyzing these data sets is similar to our previous analysis[6] Our data are blinded with a blinding factor in the range of 0.99986-1.00171 that hides the actual holding time from analyzers, as described in Ref. 6. We unblinded our data after achieving agreement between at least two analyzers for each of the three data sets.

Subsequent to unblinding, we discovered an issue that arose from an early fit to a small number of runs in the 2020 data set that resulted in a bias that caused the analyzers to exclude seven runs because of high $X^2$. After unblinding, these runs were found to have a $X^2$ below our cutoff threshold of 10, and they were added back into the unblinded data set, resulting a 1.0 s shift to the fitted lifetime for a portion the 2020 data set (659 runs), which compares to the 1.0 s statistical uncertainty associated with this subset of data.

Data were taken in octets of runs, with different holding times chosen to minimize impacts of drifts in normalization (20,1550,1550,50,100,1550,1550,200 s). The number of short and long (1550 s) holding runs was chosen to optimize precision in determining the lifetime. In practice, the number of runs with 1550 s holding time was roughly equal to the total number of the shorter holding time runs.

The elements of the analysis are run selection, event definition, correction for rate dependent effects, calculating a yield normalized to the number of trappable neutrons loaded into the trap, determining a lifetime from the long and short yields, and finally applying systematic corrections to the lifetime.

### Event Definition

The data stream consists of ordered lists of time stamps and channel numbers for the monitors and dagger photomultiplier tubes. The monitor detector signals were integrated with timing-filter amplifiers and discriminated, such that a detector hit can be interpreted as a single UCN-absorption event. In the case of the dagger detector, the times of single photoelectrons from each PMT are recorded. The data stream also includes a map of the state of a set of tag bits that mark the state of experimental controls with each time stamp. A UCN event in the dagger detector is defined by time clustered-photon coincidences between pairs of dagger phototubes. We have used different approaches to event definition and tested the impact by analysis of pseudo data to inform our choice.

The search for an event starts with observation of two photons (from different phototubes) within 100 ns. The photon number is incremented with each additional photon until the time between subsequent photons is greater than 1000 ns. A 20 ns fixed deadtime is built into the event detection to avoid retriggering. The standard for determining a UCN event requires a minimum number of photons during the clustering period. The features above (number of photons, coincidence windows, fixed deadtime) can be varied. The live time during the complete measurement cycle (fill, clean, and unload) is tracked in a time-binned histogram by accumulating the dead time from each UCN event clustering period. We tested various UCN event threshold photon numbers and converged on using ten photons, as it improved signal-to-noise for long holding times with only slight degradation in statistical uncertainty.

Corrections are made for the rate-dependent effects of deadtime and photon pileup. At the finish of analyzing a run, the number of coincidence events in each time bin is divided by the livetime for that bin, and the resulting histogram is output as our "unloads" versus time. The data are also corrected for extra photons that are the result of previous events that depend on the UCN rate, the pileup correction. The pileup correction is performed using two methods. First, by tracking the instantaneous photon rate, excluding photons within UCN events, and adjusting the photon threshold used to define UCN events to account for the estimated probability of accidental photons within an event. Second, by using a model of the photon tail to statistically correct for pileup photons.

A comparison of dead time and pile up corrections for the UCN counting portion of an average of 20 s holding time runs for the different dagger configurations is shown in

Figure 5. These corrections scale with counting rate.

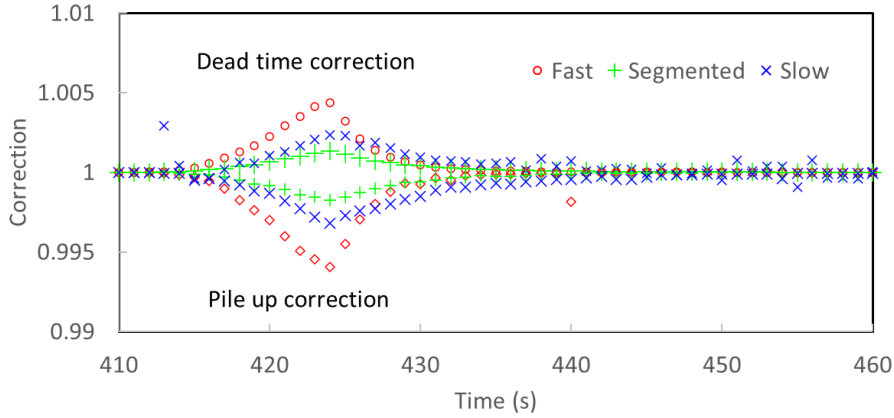

Figure 5) Example of 20 s holding time deadtime and pileup correction for the average short holding runs as shown in Figure 4) .

*Run selection*

"Good" runs were chosen from a large set of production runs. Runs were divided into groups (called epochs) for separate analysis for reasons such as a major change in experimental configuration or a change in blinding factors. It was not infrequent for the proton beam used to generate the UCN to be interrupted during trap filling. Runs are eliminated when this led to significant changes in monitor counter ratios or a very small number of loaded UCN. Off-normal conditions in the experiment noted in the logbook also resulted in elimination of runs. Runs that exhibit yields far from others with the same storage time are examined for potential problems and eliminated if warranted. Each analyzer developed and shared their selection criteria. In the end about 90% of the production runs pass the run selection criteria. This number varies for the different analyzers because of different selection criteria. The lifetimes extracted by the different analyzers agreed to within the uncorrelated errors between their data sets.

*Calculation of Yields*

The analysis uses a histogram of UCN detected (as defined in the Event Definition section above) in the dagger detector as a function of time. The histogram has a well-defined peak at the time the dagger is lowered to empty the trap. UCN are integrated over the peak region (usually 60 s) and a background region of the same length starting 50 s later in time. For a ten-photon coincidence event, the integrated peak/background ratio is typically greater than 150 for a 1550 s holding time run. Unloading time distribution for different holding times are shown in Figure 6.

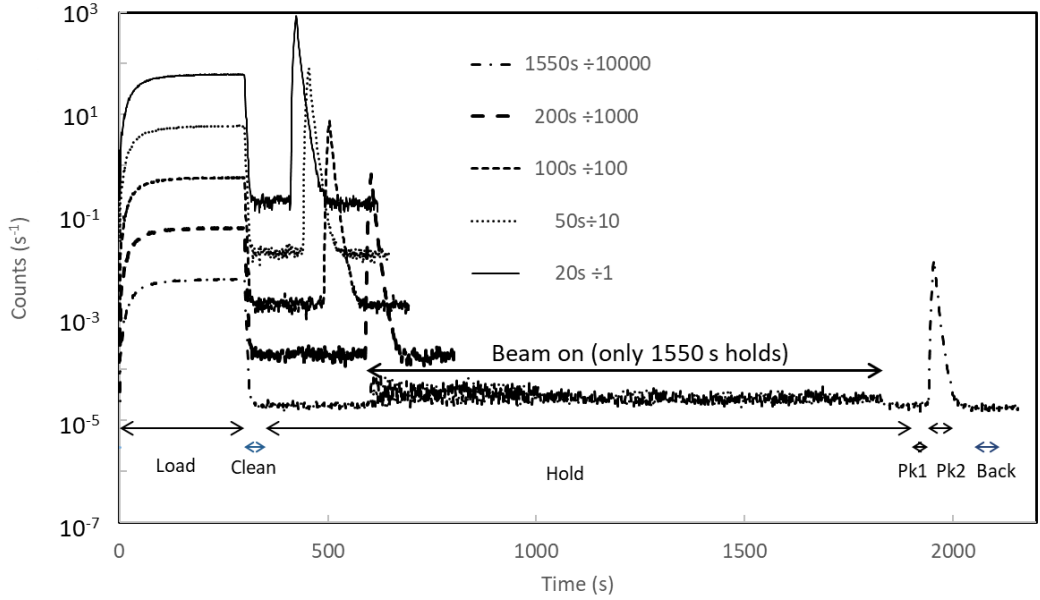

Figure 6) Average holding time distributions from the 2022 running period. During about half of the long holding time runs, the proton beam was turned on to supply UCN to other experiments. This is reflected in the larger backgrounds labeled in the figure.

Background is subtracted from the peak, and yields are calculated relative to a monitor (the RH Dump detector described earlier) that measures a quantity proportional to the number of UCN loaded into the trap. Those UCN remaining in the RH at the end of the fill are emptied into the RH Dump detector. Yields, $Y_{0,i}$, and uncertainties, $\Delta Y_{0,i}$, for a given run, $i$, are calculated as follows:

$$Y_{0,i} = \frac{C_{peak,i} - C_{back,i}}{M_i}$$

$$\Delta Y_{0,i} = \frac{1}{M_i} sqrt\left[ \frac{\left(C_{peak,i} + C_{back,i}\right)}{DQE_{peak}^2} + \frac{\left(C_{peak,i} - C_{back,i}\right)^2}{M_i \cdot DQE_{Norm}^2} \right] \qquad (2)$$

where $C_{peak,i}$ is the sum of UCN counts in a counting gate beginning at the time the dagger is lowered into its counting position, $C_{back,i}$ is the background obtained in the background gate, and $M_i$, is the integrated number of counts measured in the RH Dump Detector at the end of the filling period. The factor $DQE_{peak}$ and $DQE_{norm}$ is adjusted in the final lifetime fir to give a reduced $X^2$ of unity in the lifetime fit (discussed more below). In this work, we have assumed that fluctuations in the loading result in yield fluctuations that are not reflected in the uncertainty in $M_i$, which is several time $10^5$ and thus has a small uncertainty, $DQE_{norm}$ has been used for this adjustment and has been adjusted using the short holding time runs. The factor $DQE_{peak}$ is adjusted in the final lifetime fits to achieve $X^2$ of unity. In our previous work, this adjustment was made using only $DQE_{norm}$. The differences are small.

The normalization is based on a different UCN velocity spectrum than is stored in the trap after cleaning. We apply a global correction for this effect over sets of runs (an "epoch") for which the spectrum is nearly constant using a ratio ($R_{mon,i}$) of the RH monitor to a second monitor more sensitive to the narrow energy range neutrons that are stored in the trap. Data with the roundhouse cleaner removed and with a dam in roundhouse to cut off the low energy part of the spectrum have been used provide a check of this correction, which has the form:

$$Y_{sc,i} = Y_{0,i}\left(1 + S_i\left(R_{mon,i} - \langle R_{mon,i}\rangle\right)\right), \tag{3}$$

where $Y_{0,i}$ is corrected by a small number ($S_i$) for each run relative to the average of $R_{mon,i}$ over an epoch. The $S_i$ corrections are chosen to minimize the $\chi^2$ between $Y_{sc,i}$ and a calculated yield $Y_{cal,i}$ for the short (<1550 s) holding time runs in an epoch.

$Y_{cal,i}$ is the yield predicted for a given run, $i$, for a given storage time, $T_{hold,i}$, and lifetime, $\tau_{fit}$:

$$Y_{cal,i} = e^{-\left(\frac{T_{hold,i} + \Delta T_{ps,i}}{\tau_{fit} - \Delta \tau_{pressure,i} - \Delta \tau_{sb,i}}\right)} \tag{4}$$

We apply correction factors to the lifetime to account for phase space evolution, $\Delta T_{ps,i}$, of the UCN population in the trap the residual gas pressure in the trap, $\Delta \tau_{pressure,i}$ (which can cause UCN upscattering) and the statistical bias, $\Delta \tau_{sb,i}$, that arises from combining many measurements that follow Poisson statistics. These factors are discussed in the next section.

Degradation of the surface of the solid deuterium (SD$_2$) of the UCN source results in both reduced output and hardening of the UCN spectrum[48]. We carried out a warming and refreezing of the SD$_2$ multiple times during each running period to keep the source production as high as possible. These time-dependent normalization changes were accounted for with a second averaging using $Y_{cal,i}$ to smooth out intermediate, smooth variations (as opposed to step changes) in the monitors. We use an average of the ratio of ($Y_{SC,i}/Y_{cal,i}$) for runs in a continuous block of $n$ runs, using an average of that ratio for the subset of the $n$ runs with storage time less than 1550 s (typically $n$ is 15, and half the runs in that block of 15 will be short holding time runs). This leads to a correction factor, $CF_i$, and corrected yields, $Y_{cor,i}$:

$$CF_i = \left\langle \frac{Y_{sc,i}}{Y_{cal,i}} \right\rangle_n \tag{5}$$

$$Y_{cor,i} = \frac{Y_{sc,i}}{CF_i}. \tag{6}$$

This procedure removes any remaining long-term drifts without the need for complete octets. Excluding the long holding time runs from the calculation of CF reduces any possible correlation between the correction factor and the fitted lifetime. This procedure leads to a larger useful data set because accelerator failures in the early filling stage, leading to excluded bad runs, were nonnegligible. The uncertainty, $\Delta Y_{cor,i}$, has been propagated from the result of Equation 2.

The lifetime is fitted to the ensemble $Y_{corr,i}$ in the data set. The $Y_{corr,i}$ are assumed to be normally distributed. The value of the lifetime is calculated by minimizing $\chi^2$ of the set. The factor, $DQE_{peak}$, is applied to account for unknown efficiency effects and quality of the fit. The goal is to achieve a reduced $\chi^2$ of 1 and then calculate the statistical uncertainty in $\tau_{fit}$ by changing $\tau$ to change the overall $\chi^2$ by 1.

$$\chi^2 = \sum_i \left( \frac{Y_{corr,i} - Y_{cal,i}}{dY_{corr,i}} \right)^2 \tag{7}$$

*Measured Systematic Corrections to the Lifetime*

Systematic corrections are determined from the data for phase space evolution, residual pressure in the trap, and statistical bias introduced when combining runs with different statistical accuracies. These corrections modify the lifetime calculation. The lifetime is fitted with each correction on and off to determine the size of each these corrections.

The phase space evolution, $\Delta T_{ps}$, is a correction to the storage time of the long holding time runs due to the evolution of the UCN phase space distribution in the trap, which changes the average time it takes a UCN to encounter the lowered dagger. This is determined by calculating the average of the detection times of the unload counts relative to the time the dagger is lowered, over all runs of a given storage time, and correcting the storage time used in the calculation of yield for the long holding time based on the average difference between expected peak position of 1550 s and 20 s runs:

$$\Delta T_{ps,i} = \left( \left\langle X_{1550} \right\rangle - \left\langle X_{20} \right\rangle \right) \tag{8}$$

This shift is applied to each of the 1550 s holding time runs as shown in Equation 4. This shift is less than 0.01 s with a negligible uncertainty.

The pressure correction, $\Delta\tau_{pressure,i}$ is based on published cross section measurements[49, 50]. The correction uses the measured pressure ($P$) in the trap for each run:

$$\Delta\tau_{pressure} = 3.38 \times 10^5 \frac{s}{mbar} P(mbar) \tag{9}$$

and the assumption that the gas in trap is water vapor. This correction is applied to each run based on the trap pressure for that run.

The final correction is based on the well-known effect in averages resulting from many data sets that are Poisson distributed. Smaller numbers have smaller absolute errors and are weighted more heavily in a weighted average. This lowers the long holding time yield relative to the short holding time yield which results in a systematically shorter extracted lifetime. This is the statistical bias correction, $\Delta\tau_{SB}$. The correction is applied in the calculated lifetime as shown in Equation 4 above. The value of $\Delta\tau_{SB}$ is determined by a model data set consisting of a number (192 octets, 1536 runs) of octets (described above). For each octet, a Poisson-distributed random number of counts was chosen from a distribution with standard deviation of $\sqrt{N}$ and a mean given by:

$$N = N_0 e^{-\frac{t_{hold}}{\tau_{mc}}} \tag{10}$$

The octets were analyzed using the same procedure as with the production runs to obtain a fitted lifetime. This includes renormalizing each octet using ratios to the predicted lifetime. The entire simulated data set is then fitted to obtain a biased lifetime, $\tau_{biased}$.

This procedure is repeated 10000 times and the fitted average lifetime is subtracted from the lifetime assumed in the Monte Carlo, $\tau_{mc}$=877.75 s, to obtain the statistical bias. The uncertainty in $\tau_{fit}$ was calculated using the standard deviation of the result over the 10000 trials.

This procedure is repeated to generate results for a range of $N_0$. The difference between the Monte Carlo lifetime and the fitted biased lifetime is fitted as a function of $N_0$ using a power law:

$$\Delta\tau_{sb} = aN_0^{\left(b + \frac{c}{N_0}\right)} \tag{11}$$

The parameters used for *a, b,* and *c* are (2985.3, -1.00102, and 0.855). When the performance of the UCN source improved, $N_0$ increased, decreasing the correction. In practice, different functional fits to the statistical bias Monte Carlo result in equivalent statistical bias corrections. The *c* term in the power law is needed to produce good agreement for the smallest values of $N_0$ (<1000), which are only seen when vertical sections in the 2022 dagger are analyzed separately. For the data presented in this paper, the statistical bias correction was between 0.4 and 0.5 s.

## UCN Heating and Cooling

Quasi-bound UCN in quasi periodic orbits can remain in the trap for long times relative to the long holding time before escaping. UCN that are vibrationally heated can also escape the trap at later times. Both of these processes, discussed in Reference 51, can lead to trap lifetimes that differ from the free neutron lifetime. These effects can both be experimentally limited using the first counting period (peak 1) when the dagger is moved to the cleaning position before being lowered to the bottom of the trap.

We have taken data without lowering the cleaner ("uncleaned") to determine the losses experienced when UCN are not properly cleaned, as in Reference 6, The measured lifetime for uncleaned runs is significantly shorter due to UCN above the cleaning height. The observed losses are scaled for the relative magnitude of the peak 1 signal to account for the significant difference in populations above the cleaning height determining the systematic uncertainty for uncleaned UCN.

Unloading time distributions for 20 s holding time runs without cleaning (i.e., in which the cleaner was never lowered into the trap prior to the holding period) are shown in Figure 7. The peak of high energy UCN remaining in the uncleaned distributions (peak 1) show a tail that extends under the normal second counting period (peak 2). The fraction of counted neutrons has been estimated by taking data with a longer peak 1 counting time of 200 s. The ratio of background subtracted counts in first 200 s of peak 1 to those in the first 40 s (the duration the dagger is held in this position in production runs) is 1.76(0.15) and has been used to correct the net counts in peak 1 in the production data to arrive at the corrections and uncertainties for inefficient cleaning and heating given in Table 1.

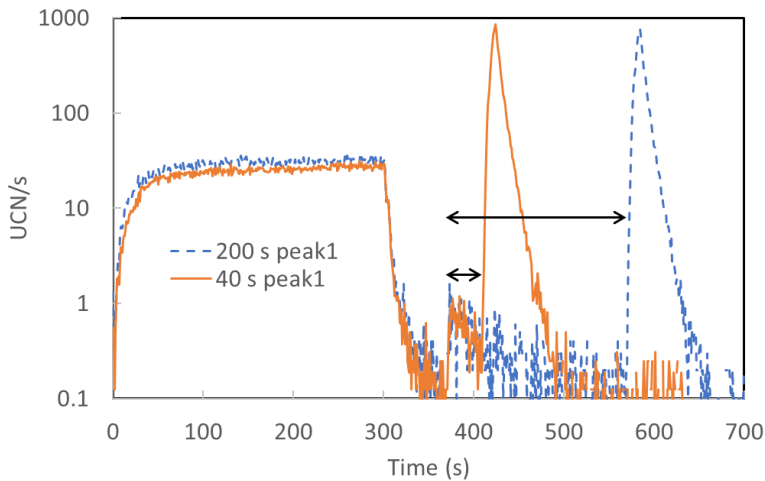

Figure 7) Average counting time distribution for 2023 short holding time runs with no cleaning with a 40 s long peak 1 counting time (red solid line) and with 200 s long peak 1 counting time (blue dashed lne). The arrows show the integration gates used for this analysis.

## Impacts of Phase Space Evolution and Dagger Non-uniformity

The UCN$\tau$ trap is designed with a built-in asymmetry to mix the neutron orbits to rapidly populate all the available phase space in the trap[52]. However, detailed Monte Carlo studies[51] have shown that the process of filling the phase space is somewhat slower than originally expected. For example, neutrons loaded into the trap must have vertical momentum to move into the trap, clearly leaving some parts of phase space initially unoccupied. Over time, neutrons evolve into the unoccupied phase space.

In our previous work, we accounted for phase space evolution by using the measured mean UCN unload time for long holding time runs (see Equation 8) rather than the programmed unload time in the lifetime fitting. Although this requires a correction to the lifetime, it introduces a negligible systematic uncertainty.

In the 2022 data from this work, the segmented dagger (Figure 3) provided additional information about phase space evolution. We measured independent lifetimes for each of the dagger segments and found that the fitted lifetime across the dagger segments varies by about 10 s (Figure 8). The gains of the sections of the segmented dagger, measured using the distribution of the number of photoelectrons from each section, varied by 20%. Balancing the gains reduced the fitted lifetime by 0.22 s. The lifetime reported here uses balanced gains for the 2022 data set.

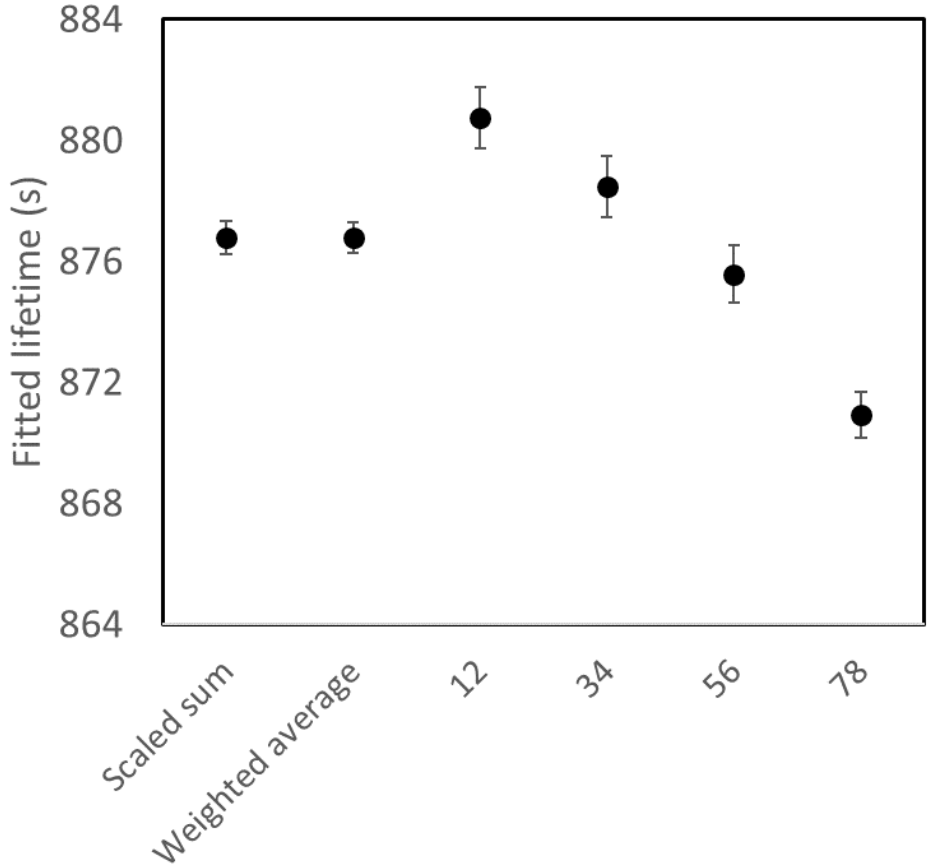

Figure 8) 2022 lifetime fits comparing the results from the 4 different strips with the analysis of the scaled sum and weighted average. The four strips (12, 34, 56, 78) are shown from left to right in Figure 3.

The lifetime found from the shortest strip (78), which does not reach as far into the trap as the longer strips, is the shortest. We interpret this as due to phase space mixing, which causes the distribution of the neutrons in the trap to move over time. This interpretation is supported by ongoing Monte Carlo modelling[51].

There is only a small shift in the ratio of the long to short holding time unloaded counts as a function of time as the dagger is moved into the trap, Figure 9. Similar plots for the individual dagger strips show similar ratios vs time. This indicates that there is very little vertical phase space evolution. A change in the vertical distribution between long and short holding time would be reflected in a shift in the centroid of the time distributions which is measured to be less than 0.04 s. These shifts have been included in the lifetime fits as a shift in the holding time.

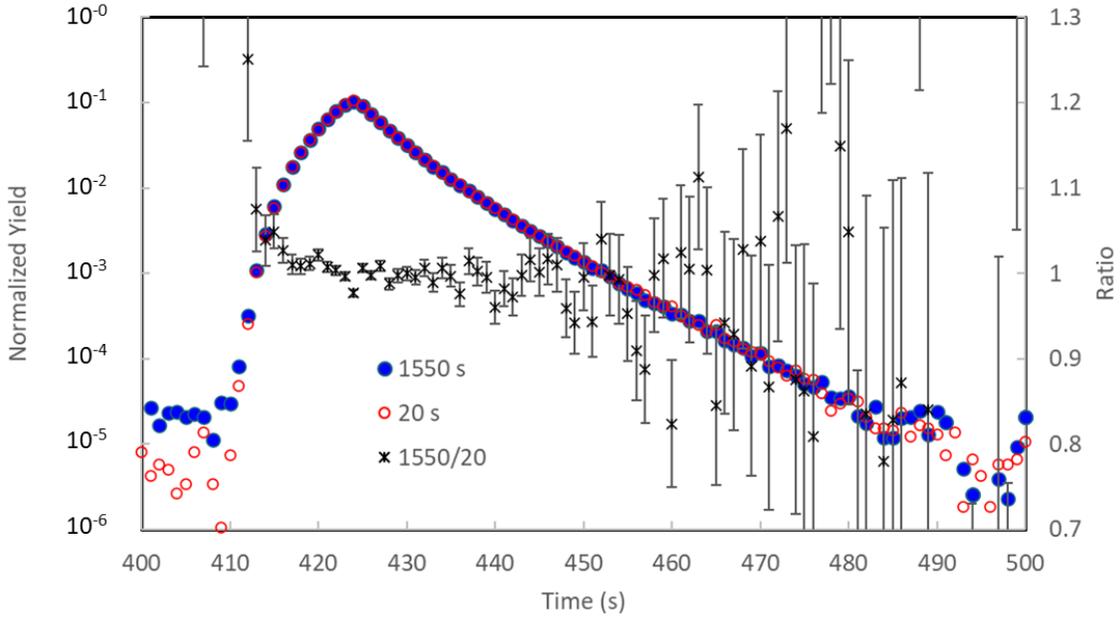

Figure 9) Normalized short (red open) and long (blue closed) unloading curves plotted on the left axis and the ratio, plotted on the right axis.

If all the neutrons are counted, phase space evolution would not affect the measured lifetime.  However, if different parts of the dagger have different efficiencies that bias the counting as a function of position, they will bias the lifetime determination. To estimate the size of this systematic uncertainty, the light output from the daggers was mapped using a laser to excite the wavelength shifting fibers. The maps show that the relative dagger efficiency has top to bottom efficiency changes of less than 3%. We changed the weighting of the different dagger segments (Figure 10) and refitted the lifetime to estimate the uncertainty introduced by non-uniformities in the horizontal dagger efficiency.

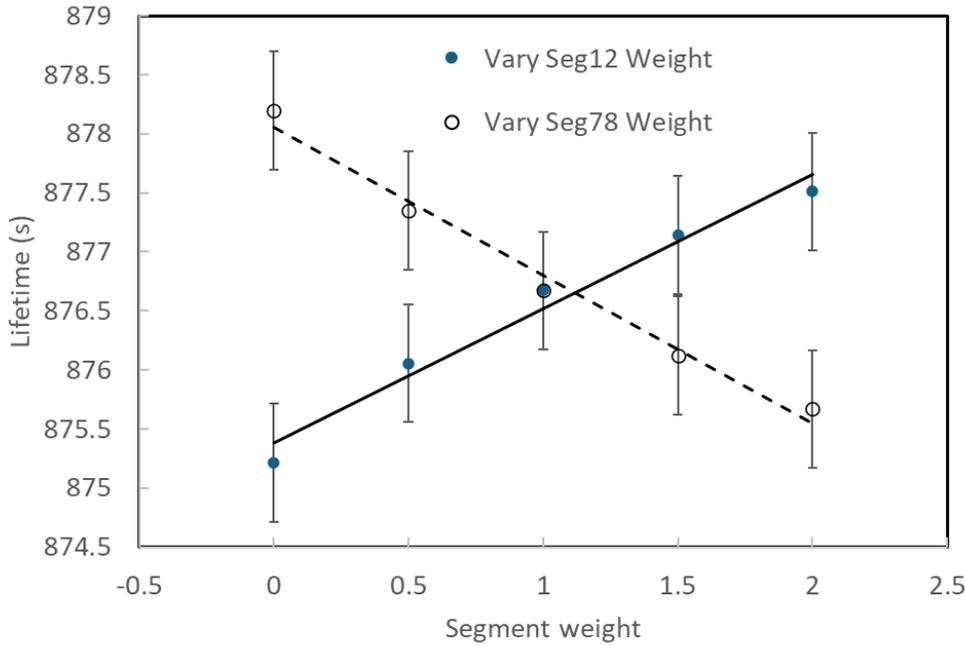

Figure 10) A plot of the lifetime is a function of changes in the weighting of the two outer segments. The lines show the slopes of the two curves used to estimate the horizontal phase space evolution uncertainty.

We find that these gain variations introduce a systematic uncertainty of 0.02 s (listed in Table 1).

Table 1) Systematics corrections and uncertainties calculated for each year's data set. Numbers shown are an averages of the different analyzers' individual corrections and uncertainties. The "heating" and "uncleaned" corrections are a weighted average of those in this work and reference 6. All numbers are given in seconds.

| Effect | 2020 | | 2021 Fast | | 2021 Slow | | 2022 | | Average | |
|---|---|---|---|---|---|---|---|---|---|---|
| | Correction | Uncertainty | Correction | Uncertainty | Correction | Uncertainty | Correction | Uncertainty | Correction | Uncertainty |
| Event definition | 0.00 | 0.20 | 0.00 | 0.16 | 0.00 | 0.17 | 0.00 | 0.14 | 0.00 | ±0.16 |
| Dagger uniformity | 0.00 | 0.02 | 0.00 | 0.02 | 0.00 | 0.02 | -0.22 | 0.02 | 0.06 | ±0.02 |
| Residual gas | 0.06 | 0.03 | 0.07 | 0.04 | 0.04 | 0.02 | 0.05 | 0.03 | 0.05 | ±0.03 |
| Statistical bias | 0.51 | 0.01 | 0.50 | 0.01 | 0.55 | 0.01 | 0.31 | 0.01 | 0.47 | ±0.01 |
| Depolarization | 0.00 | 0.07 | 0.00 | 0.07 | 0.00 | 0.07 | 0.00 | 0.07 | 0.00 | +0.07 |
| Uncleaned | 0.00 | 0.01 | 0.00 | 0.01 | 0.00 | 0.01 | 0.00 | 0.01 | 0.00 | +0.01 |
| Heating | 0.00 | 0.07 | 0.00 | 0.07 | 0.00 | 0.07 | 0.00 | 0.07 | 0.00 | +0.07 |
| $\Delta t_{ph}$ | -0.02 | 0.02 | -0.01 | 0.01 | -0.02 | 0.01 | -0.04 | 0.01 | -0.02 | ±0.01 |
| Uncorrelated sum | | | | | | | | | 0.58 | +0.20-0.17 |

## RESULTS AND CONCLUSION

Four different analysis teams prepared results for this publication. Each team worked on data from various years, and we ensured there would be at least three independent analyses which agree within the uncorrelated error for each year. The lifetime results for each year are the average of all the independent results ($\tau_A$, $\tau_B$, $\tau_C$, and $\tau_D$) for that year, with a statistical uncertainty chosen as the average of the uncertainties estimated by individual analyzers ($d\tau_A$, $d\tau_B$, $d\tau_C$, and $d\tau_D$). The combined lifetime in Table 2 is the error-weighted average of each year's results with uncertainty given by weighted standard mean.

Table 2 Unblinded neutron lifetime results by different analysis teams (A, B, C, and D) for each year and the combined result. The previous published results from Reference [6] are included in the global average.

| Analysis | A | B | C | D | Average |
|---|---|---|---|---|---|
| 2018 ref 6 | 877.68±0.30 | 877.78±0.34 | 877.74±0.33 | | 877.73±0.32 |
| 2019 ref 6 | 878.06±0.49 | 877.80±0.46 | 877.55±0.55 | | 877.80±0.50 |
| 2020 | 879.47±0.86 | 879.38±0.92 | 879.32±0.90 | | 879.39±0.89 |
| 2021 | 878.41±0.60 | 878.40±0.59 | 878.41±0.55 | | 878.41±0.58 |
| 2022 | 876.78±0.53 | 876.78±0.58 | 877.06±0.53 | 877.08±0.63 | 876.93±0.57 |
| Current | | | | | 877.95±0.37 |
| Global | | | | | 877.82±0.22 |

Each team also studied various systematic corrections mentioned in the Analysis section earlier, and the size of each correction is reported in Table 1. Table 1 also shows the average of the systematic uncertainties estimated by each team for each year. The statistical-bias correction (~0.5 s) is the largest systematic correction. The table also reports the average of systematic uncertainties stemming from each team's event definition parameters, and it is the largest contributor to the systemic uncertainty in the reported lifetime, followed by the systematic uncertainty from the residual uncleaned or heated UCNs in the trap. Varying the event definition parameters (event length, counting length, and threshold) yields a distribution of lifetimes; the uncertainty reported is one-sigma of that distribution.

## DISCUSSION

The results of this analysis have been included in a global analysis that includes data from the previous two UCNτ measurements[6]. The result, plotted in Figure 11, shows good agreement with each individual analysis. The p-value comparing the combined result with the individual year results is 14%. Our result for the free neutron lifetime measured with trapdoor loading is 877.82±0.22 (statistical)+0.20-0.17 (systematic) s. The systematic uncertainty is taken from the quadratic sum of the average values reported in Table 1. We have taken the systematic uncertainties from the current analysis to apply to the previous data.

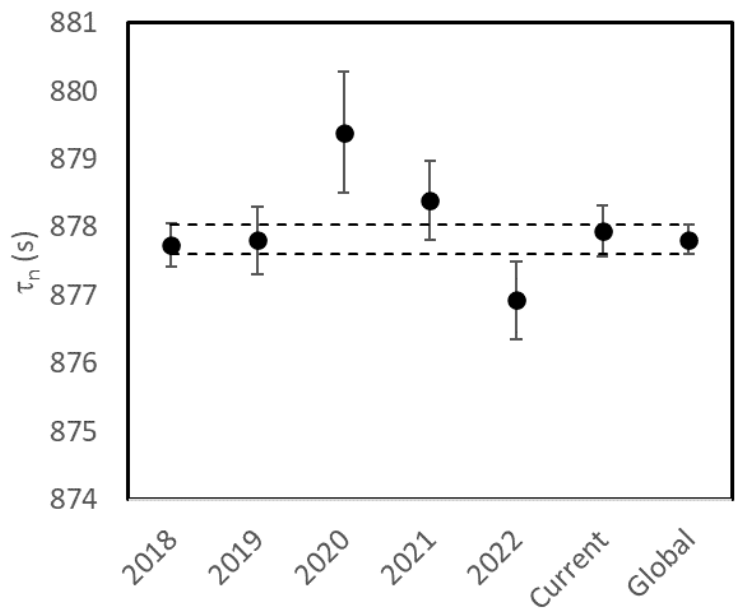

Figure 11) The global analysis (horizontal lines) compared to the individual year's results for all of the included data.

Planned upgrades to the UCNτ apparatus: UCNτ+

This paper presents the final lifetime from the Los Alamos UCN facility magneto-gravitational trap using loading through the trap door, UCNτ. The result is $\tau_n$=877.82±0.22 (statistical)+0.20-0.17 (systematic) s. With the current counting techniques and given the number of UCN loaded into the experiment through the trap door, it is difficult to obtain better statistical or lower systematic uncertainties. Work is underway to replace the trapdoor-based loading with an elevator loading method that is expected to improve the loaded number of UCN by a factor of 5-10. The largest systematic uncertainty arises from the event definition. We are currently developing a new screen scintillator based on cerium doped yttrium aluminum perovskite (YAP:Ce), that, with a suitable wavelength shifter, provides similar light output to ZnS:Ag, but without the long fluorescence tail.  Events will be defined from coincidence of light produced with the 40 ns decay time of the YAP:Ce.  This compares to the average event width of about 4 μs with the ZnS:Ag detector, and should provide at least a factor of 10 reduction in the driving systematic uncertainty in this work. With these improvements it should be possible to approach a total uncertainty of 0.10 s.

## Acknowledgments

The authors would like to thank the staff and management of the Los Alamos Neutron Science Center for providing the UCN used for these experiments and the U.S. Department of Energy, Office of Science, Office of Nuclear Physics under Awards No. DE-FG02-ER41042, No. DE-AC52-06NA25396, No. DE-AC05-00OR2272, and No. 89233218CNA000001 under proposal LANLEEDM; NSF Grants No. 1614545, No. 1914133, No. 1506459, No. 1553861, No. 2310015, No. 1812340, No. 1714461, No. 2110898, No. 1913789, and No. 2209521; and NIST precision measurements grant.

# REFERENCES

1. G. J. Mathews, T. Kajino and T. Shima, Physical Review D **71** (2), 021302 (2005).
2. D. Dubbers and M. G. Schmidt, Reviews of Modern Physics **83** (4), 1111 (2011).
3. M. Gonzalez-Alonso, O. Naviliat-Cuncic and N. Severijns, Progress in Particle and Nuclear Physics **104**, 165-223 (2019).
4. A. Ivanov, M. Pitschmann and N. Troitskaya, Physical Review D **88** (7), 073002 (2013).
5. S. Navas, C. Amsler, T. Gutsche, C. Hanhart, et al., Physical Review D **110** (3), 030001 (2024).
6. F. M. Gonzalez, E. Fries, C. Cude-Woods, T. Bailey, et al., Physical review letters **127** (16), 162501 (2021).
7. A. Saunders, M. Makela, Y. Bagdasarova, H. Back, et al., Review of Scientific Instruments **84** (1) (2013).
8. T. M. Ito, E. Adamek, N. Callahan, J. Choi, et al., Physical Review C **97** (1), 012501 (2018).
9. F. E. Wietfeldt and G. L. Greene, Reviews of Modern Physics **83** (4), 1173-1192 (2011).
10. A. Czarnecki, W. J. Marciano and A. Sirlin, Physical Review Letters **120** (20), 202002 (2018).
11. C.-Y. Seng, M. Gorchtein, H. H. Patel and M. J. Ramsey-Musolf, Physical Review Letters **121** (24), 241804 (2018).
12. Y. Aoki, T. Blum, G. Colangelo, S. Collins, et al., The European Physical Journal C **82** (10), 869 (2022).
13. A. Czarnecki, W. J. Marciano and A. Sirlin, Physical Review D **100** (7), 073008 (2019).
14. L. Hayen, Physical Review D **103** (11), 113001 (2021).
15. C.-Y. Seng, M. Gorchtein and M. J. Ramsey-Musolf, Physical Review D **100** (1), 013001 (2019).
16. M. Gorchtein and C.-Y. Seng, Universe **9** (9), 422 (2023).
17. V. Cirigliano, W. Dekens, E. Mereghetti and O. Tomalak, Physical Review D **108** (5), 053003 (2023).
18. V. Cirigliano, D. Díaz-Calderón, A. Falkowski, M. González-Alonso, et al., Journal of High Energy Physics **2022** (4), 1-61 (2022).
19. V. Cirigliano, A. Crivellin, M. Hoferichter and M. Moulson, Physics Letters B **838**, 137748 (2023).
20. V. Cirigliano, W. Dekens, J. de Vries, E. Mereghetti, et al., Journal of High Energy Physics **2024** (3), 1-69 (2024).
21. V. Cirigliano, W. Dekens, J. de Vries, S. Gandolfi, et al., arXiv preprint arXiv:2405.18469 (2024).
22. T. Aaltonen, S. Amerio, D. Amidei, A. Anastassov, et al., Science **376** (6589), 170-176 (2022).
23. B. Belfatto and S. Trifinopoulos, Physical Review D **108** (3), 035022 (2023).
24. O. Fischer, B. Mellado, S. Antusch, E. Bagnaschi, et al., The European Physical Journal C **82** (8), 665 (2022).
25. D. Marzocca and S. Trifinopoulos, Physical Review Letters **127** (6), 061803 (2021).
26. A. Falkowski, M. González-Alonso and O. Naviliat-Cuncic, Journal of High Energy Physics **2021** (4), 1-36 (2021).
27. J. S. Nico, M. S. Dewey, D. M. Gilliam, F. E. Wietfeldt, et al., Physical Review C **71** (5), 055502 (2005).
28. A. Yue, M. Dewey, D. Gilliam, G. Greene, et al., Physical review letters **111** (22), 222501 (2013).
29. A. Serebrov, V. Varlamov, A. Kharitonov, A. Fomin, et al., Physics Letters B **605** (1-2), 72-78 (2005).
30. A. Pichlmaier, V. Varlamov, K. Schreckenbach and P. Geltenbort, Physics Letters B **693** (3), 221-226 (2010).
31. A. Steyerl, J. Pendlebury, C. Kaufman, S. S. Malik, et al., Physical review C **85** (6), 065503 (2012).

32.     S. Arzumanov, L. Bondarenko, S. Chernyavsky, P. Geltenbort, et al., Physics Letters B **745**, 79-89 (2015).

33.     A. Serebrov, E. Kolomensky, A. Fomin, I. Krasnoshchekova, et al., Physical Review C **97** (5), 055503 (2018).

34.     V. Ezhov, A. Andreev, G. Ban, B. Bazarov, et al., JETP Letters **107** (11), 671-675 (2018).

35.     G. L. Greene and P. Geltenbort, Scientific American **314** (4), 36-41 (2016).

36.     B. Fornal and B. Grinstein, Physical review letters **120** (19), 191801 (2018).

37.     G. Baym, D. Beck, P. Geltenbort and J. Shelton, Physical review letters **121** (6), 061801 (2018).

38.     D. McKeen, A. E. Nelson, S. Reddy and D. Zhou, Physical Review Letters **121** (6), 061802 (2018).

39.     J. Ellis, G. Hütsi, K. Kannike, L. Marzola, et al., Physical Review D **97** (12), 123007 (2018).

40.     X. Sun, E. Adamek, B. Allgeier, M. Blatnik, et al., Physical Review C **97** (5), 052501 (2018).

41.     Z. Tang, M. Blatnik, L. J. Broussard, J. Choi, et al., Physical Review Letters **121** (2), 022505 (2018).

42.     M. Le Joubioux, H. Savajols, W. Mittig, X. Fléchard, et al., Physical Review Letters **132** (13), 132501 (2024).

43.     R. Pattie Jr, N. Callahan, C. Cude-Woods, E. Adamek, et al., Science **360** (6389), 627-632 (2018).

44.     D. J. Salvat, E. R. Adamek, D. Barlow, J. D. Bowman, et al., Physical Review C **89** (5), 052501 (2014).

45.     C. Morris, E. Adamek, L. Broussard, N. Callahan, et al., Review of Scientific Instruments **88** (5) (2017).

46.     Z. Wang, M. Hoffbauer, C. Morris, N. Callahan, et al., Nuclear Instruments and Methods in Physics Research Section A: Accelerators, Spectrometers, Detectors and Associated Equipment **798**, 30-35 (2015).

47.     C. Morris, E. Adamek, L. Broussard, N. Callahan, et al., Review of Scientific Instruments **88** (5), 053508 (2017).

48.     A. Anghel, T. Bailey, G. Bison, B. Blau, et al., The European Physical Journal A **54** (9), 1-15 (2018).

49.     S. Seestrom, E. Adamek, D. Barlow, L. Broussard, et al., Physical Review C **92** (6), 065501 (2015).

50.     S. J. Seestrom, E. R. Adamek, D. Barlow, M. Blatnik, et al., Physical Review C **95** (1), 015501 (2017).

51.     N. Callahan, C.-Y. Liu, F. Gonzalez, E. Adamek, et al., Physical Review C **100** (1), 015501 (2019).

52.     P. Walstrom, J. Bowman, S. Penttila, C. Morris, et al., Nuclear Instruments and Methods in Physics Research Section A: Accelerators, Spectrometers, Detectors and Associated Equipment **599** (1), 82-92 (2009).